\documentclass[runningheads]{llncs}
\usepackage{graphicx}
\usepackage{booktabs}
\usepackage{amsfonts}
\usepackage{amsmath}
\usepackage{tikz}
\usetikzlibrary{shapes,snakes}
\begin{document}
\title{Decision Trees with Hypotheses for Recognition of Monotone Boolean Functions and for Sorting}
\titlerunning{Decision Trees with Hypotheses for Two Problems}
%
\author{Mohammad Azad\inst{1}\orcidID{0000-0001-9851-1420} \and
Igor Chikalov\inst{2}\orcidID{0000-0002-1010-6605} \and
Shahid Hussain\inst{3}\orcidID{0000-0002-1698-2809} \and
Mikhail Moshkov\inst{4}\orcidID{0000-0003-0085-9483} \and
Beata Zielosko\inst{5}\orcidID{0000-0003-3788-1094}}
\authorrunning{M. Azad et al.}
%
\institute{Department of Computer Science, College of Computer and Information Sciences, Jouf University, Sakaka~72441, Saudi Arabia\\
\email{mmazad@ju.edu.sa}
\and
Intel Corporation, 5000 W Chandler Blvd, Chandler, AZ 85226, USA\\
\email{igor.chikalov@gmail.com}
\and
Department of Computer Science, School of Mathematics and Computer Science, Institute of Business Administration, University Road, Karachi 75270, Pakistan\\
\email{shahidhussain@iba.edu.pk}
\and
Computer, Electrical and Mathematical Sciences \& Engineering Division and Computational Bioscience Research Center, King Abdullah University of Science and Technology, Thuwal 23955-6900, Saudi Arabia\\
\email{mikhail.moshkov@kaust.edu.sa}
\and
Institute of Computer Science, Faculty of Science and Technology, University of Silesia in Katowice, B\c{e}dzi\'{n}ska 39, 41-200 Sosnowiec, Poland\\
\email{beata.zielosko@us.edu.pl}
}
\maketitle              
\begin{abstract}
In this paper, we consider decision trees that use  both  queries based on one attribute each and queries based on hypotheses about values of all attributes.  Such decision trees are similar to ones studied in exact learning, where  not only membership but also equivalence queries are allowed. We investigate the problem of recognition of monotone Boolean functions with $n$ variables, $n=2, \ldots, 4$, and the problem of sorting $n$ pairwise different elements from linearly ordered set, $n=3, \ldots, 6$. For each of these problems, we compare the complexity of different types of optimal (relative to the depth or the number of realizable nodes) decision trees with hypotheses. We also study the  complexity of decision trees constructed by entropy-based greedy algorithm and analyze the length of decision rules derived from these trees.

\keywords{Decision tree \and Decision rule  \and Hypothesis  \and Recognition  \and Sorting.}
\end{abstract}

\section{Introduction}
\label{S6.0}

In contrast to exact learning \cite{Angluin88,Angluin04}, where both membership and equivalence queries are used in algorithms, conventional decision trees \cite{CART,Moshkov05,Rokach} use only attributes, which are similar to membership queries. In \cite{ent,ele,CSP,ent1}, we studied decision trees with hypotheses that in addition to attributes can use hypotheses about values of all attributes. Queries based on hypotheses can be considered as an analog of equivalence queries from exact learning. We studied five types of decision trees based on different combinations of attributes, hypotheses, and proper hypotheses -- an analog of proper equivalence queries from exact learning.

In \cite{ele}, we proposed dynamic programming algorithms for computation of the minimum complexity for decision trees of five types. In \cite{ent1}, we modified some algorithms from \cite{ele} such that they can not only find the minimum complexity of decision trees but also construct decision trees with the minimum complexity. In \cite{ent1}, we also studied the length and coverage of decision rules derived from optimal decision trees. Various experimental results obtained in \cite{ele,ent1} show that the decision trees with hypotheses can have less complexity than the conventional decision trees and that the decision rules derived from the former trees can have better length and coverage than the decision rules derived from the latter trees. These results open up some prospects for using decision trees with hypotheses as a means for knowledge representation.
Unfortunately, the dynamic programming algorithms are too time consuming. To avoid this limitation, we proposed in \cite{ent} an entropy-based greedy algorithm for the construction of decision trees of five types.

In this paper, to illustrate the use of the tools created in \cite{ent,ele,ent1}, we investigate two problems of independent interest: the problem of recognition of monotone Boolean functions with $n$ variables, $n=2, \ldots, 4$, and the problem of sorting $n$ pairwise different elements from linearly ordered set, $n=3, \ldots, 6$. Note that these problems can be reformulated as exact learning problems \cite{Angluin04}.
For each of these problems, we study the depth and the number of realizable nodes for five types of optimal (relative to the depth or the number of realizable nodes) decision trees. We study the same parameters for five types of decision trees constructed by entropy-based greedy algorithm. We also analyze the length of decision rules derived from these trees.

The main goal of this paper is to compare  decision trees with hypotheses and conventional decision trees. The obtained experimental results show that, for both problems, in almost each case we can find decision trees with hypotheses  such that the results for these trees outperform the results for conventional decision trees. The only exception is the minimum number of realizable nodes for $n=3,4$ for the first problem (see Table \ref{tab6.4a}) and for $n=5,6$ for the second problem (see Table \ref{tab6.4}).

For the problem of  recognition of monotone Boolean functions, the results for the minimum depth of conventional decision trees follow from well known result of Hansel \cite{Hansel}: the minimum depth of a conventional decision tree for the recognition of monotone Boolean functions with $n$ variables is equal to $\binom{n}{\lfloor n/2 \rfloor}+\binom{n}{\lfloor n/2 \rfloor +1}$. It is easy to show that the minimum number of nodes in a conventional decision tree for the recognition of monotone Boolean functions with $n$ variables is equal to $2M(n)-1$, where $M(n)$ is the number of monotone Boolean functions with $n$ variables. Other results are new.
For the problem of sorting, the results for the optimal conventional decision trees  are well known -- see details in Sect. 6.1 of the book \cite{book19}. Other results are new.

This paper is an essential extension of the conference paper \cite{CSP}: for the problem of sorting, we added results related to the use of greedy algorithm. The problem of recognition of monotone Boolean functions was not considered in \cite{CSP}.

The rest of the paper is organized as follows. In Sect. \ref{S6.1}, we discuss basic definitions and notation. In Sect. \ref{S6.2}, we consider the problem of recognition of monotone Boolean functions, and in Sect. \ref{S6.3} -- the problem of sorting. Section \ref{S6.3} contains short conclusions.

\section{Basic Notions and Notation}
\label{S6.1}

Let $T$ be a decision table with $n$ conditional attributes $f_{1},\ldots
,f_{n}$ that have values from the set $\omega =\{0,1,2,\ldots \}$. Rows of this table
are pairwise different and each row is labeled with a decision.
For a given row of $T$, we should recognize the decision attached to it.
To this end, we will use decision trees based on two types of queries.
We can ask about the value of a conditional attribute $f_{i}\in \{f_{1},\ldots ,f_{n}\}$
on the given row. As a result, obtain an answer of the kind $f_{i}=\delta $,
where $\delta$ is the number in the intersection of the given row and the
column $f_{i}$. We can also ask if a hypothesis $f_{1}=\delta _{1},\ldots
,f_{n}=\delta _{n}$ is true, where the numbers $\delta _{1},\ldots ,\delta _{n}$
belong to the columns $f_{1},\ldots ,f_{n}$, respectively. Either this
hypothesis is confirmed or we obtain a counterexample of the kind $%
f_{i}=\sigma $, where $f_{i}\in \{f_{1},\ldots ,f_{n}\}$ and $\sigma $ is a
number from the column $f_{i}$ that is different from $\delta _{i}$. We will say that
this hypothesis is proper if $(\delta _{1},\ldots ,\delta _{n})$ is a row
of the table $T$.

We study the following five types of decision trees:
1) using attributes,
2) using hypotheses,
3) using both attributes and hypotheses,
4) using proper hypotheses,
and
5) using both attributes and proper hypotheses.

As time complexity of a decision tree $\Gamma$ we consider its depth $h(\Gamma)$, which is equal to the maximum number of queries in a path from the root to a terminal node of the tree. We consider the number of realizable relative to $T$ nodes in the decision tree $\Gamma$ as its space complexity and denote it $L(T,\Gamma )$. A node is
called realizable relative to $T$ if the computation in the tree will
pass through this node for some row and the choice of counterexamples.

We study not only decision trees with hypotheses but also decision rules derived from them. For a decision tree $\Gamma$, we correspond in a natural way to each path from the root to a terminal node, a decision rule. We remove from this rule some conditions  that follow from previous ones (see details in \cite{ent1}). As a result, we obtain a set of ``reduced'' decision rules. For each row of the decision table $T$, we find the minimum length of a ``reduced'' decision rule that covers this row. We denote by $l(T, \Gamma)$ the arithmetic mean of these minimum lengths.

We will use the following notation:
 $h^{(k)}(T)$ denotes the minimum depth of a decision
tree of the type $k$ for $T$,  $k=1, \ldots , 5$,
and
$L^{(k)}(T)$ denotes the minimum number of nodes
realizable relative to $T$ in a decision tree of the type $k$ for $T$,   $k=1, \ldots , 5$.
To compute values $h^{(k)}(T)$ and $L^{(k)}(T)$, we use dynamic programming algorithms $\mathcal{A}_{h}$ and $\mathcal{A}_{L}$ described in \cite{ele}.

We proposed in \cite{ent} an entropy-based greedy algorithm $\mathcal{E}$ that, for a given nonempty decision table $T$ and $k\in \{1,\ldots ,5\}$, constructs a
decision tree $\Gamma_{\mathcal{E}}^{(k)}(T)$ of the type $k$ for the table $T$.
We will use the following notation:
$h_{\mathcal{E}}^{(k)}(T) =h(\Gamma_{\mathcal{E}}^{(k)}(T) )$,
$L_{\mathcal{E}}^{(k)}(T) =L(T,\Gamma_{\mathcal{E}}^{(k)}(T) )$, and
$l_{\mathcal{E}}^{(k)}(T) =l(T, \Gamma_{\mathcal{E}}^{(k)}(T) )$.

Note that the complete definitions of the notions mentioned in this section can be found in \cite{ent,ele,ent1}.

\section{Recognition of Monotone Boolean Functions}
\label{S6.2}

In this section, we study the problem of recognition of monotone Boolean functions with $n$ variables, $n=2, \ldots , 4$.
A Boolean function $f(x_{1},\ldots ,x_{n})$ is called monotone if, for any two tuples
$(a_{1},\ldots ,a_{n}),(b_{1},\ldots ,b_{n})\in \{0,1\}^{n}$ such that $a_{1}\leq b_{1},\ldots ,a_{n}\leq b_{n}$, the inequality
$f(a_{1},\ldots ,a_{n})\leq f(b_{1},\ldots ,b_{n})$ holds. For a given monotone Boolean function with $n$ variables, we should recognize it
using attributes each of which is the value of the considered function on some tuple from $\{0,1\}^{n}$. We denote by $M(n)$ the number of monotone Boolean functions with $n$
variables. It is known \cite{Church} that $M(2)= 6$, $M(3)= 20$, and $M(4)= 168$.

The problem of recognition of monotone Boolean functions with $n$ variables $x_1, \ldots , x_n$ can be represented as a decision table $R_n$ with $2^n$ conditional attributes $r_{a_1 \cdots a_n}$ corresponding to tuples $(a_1, \ldots ,a_n)$ from $\{0,1\}^{n}$
and $M(n)$ rows corresponding to monotone Boolean functions with $n$ variables.
For each such function $f$, the corresponding  row of $R_n$ is labeled with this function as the decision. In the intersection with the column labeled with the attribute $r_{a_1 \cdots a_n}$, the considered row has the value $f(a_1, \ldots ,a_n)$. The table $R_n$ can also be considered as a representation of an exact learning problem \cite{Angluin04}: conditional attributes form a domain, each row describes a concept, and rows form a concept class.
The decision table $R_2$ is shown in Fig. \ref{fig1}. 

\begin{figure}[h]
	\centering
$R_{2}=\;$%
\begin{tabular}{|cccc|c|}
\hline
$r_{00}$ & $r_{01}$ & $r_{10}$ & $r_{11}$ & \\
\hline
0 & 0 & 0 & 0 & $0$\\
1 & 1 & 1 & 1 & $1$\\
0 & 0 & 1 & 1 & $x_1$\\
0 & 1 & 0 & 1 & $x_2$\\
0 & 0 & 0 & 1 & $x_1 \wedge x_2$\\
0 & 1 & 1 & 1 & $x_1 \vee x_2$\\
\hline
\end{tabular}
\caption{Decision table $R_2$}
		\label{fig1}
\end{figure}

For $n=2, \ldots , 4$  and $k=1, \ldots , 5$, we find values of $h^{(k)}(R_n)$ and $L^{(k)}(R_n)$ using dynamic programming algorithms $\mathcal{A}_{h}$ and $\mathcal{A}_{L}$ described in \cite{ele} -- see results in Tables \ref{tab6.1a} and \ref{tab6.2a}. Minimum values for each $n$ are in bold.
The obtained experimental results show that the minimum depth of decision trees of  types 2--5 is noticeably  less than the minimum depth of decision trees of  type 1. The minimum number of realizable nodes in decision trees of types 3 and 5  is noticeably less than the minimum number  of realizable nodes in decision trees of  type 1. Decision trees of  types 2 and 4 have too many nodes.

\begin{table}[h!]
    \caption{Experimental results for dynamic programming algorithm $\mathcal{A}_{h}$}
\label{tab6.1a}
\begin{tabular}{cccccc}
\toprule
$n$  &  $h^{(1)}(R_n)$ & $h^{(2)}(R_n)$ & $h^{(3)}(R_n)$ & $h^{(4)}(R_n)$ & $h^{(5)}(R_n)$ \\ \midrule
2 & 3 & \bf 2 & \bf 2 & \bf 2 & \bf 2 \\
3 & 6 & \bf 3 & \bf 3 & \bf 3 & \bf 3 \\
4 & 10 & \bf 6 & \bf 6 & \bf 6 & \bf 6 \\
 \bottomrule
\end{tabular}
\end{table}

\begin{table}[h!]
    \caption{Experimental results for dynamic programming algorithm $\mathcal{A}_{L}$}
\label{tab6.2a}
\begin{tabular}{cccccc}
\toprule
$n$  &  $L^{(1)}(R_n)$ & $L^{(2)}(R_n)$ & $L^{(3)}(R_n)$ & $L^{(4)}(R_n)$ & $L^{(5)}(R_n)$ \\ \midrule
2 & 11 & 12 & \bf 9 & 12 & \bf 9 \\
3 & 39 & 76 & \bf 33 & 76 & \bf 33 \\
4 & 335 & 8,808 & \bf 283 & 8,808 & \bf 283 \\
 \bottomrule
\end{tabular}
\end{table}

For $n=2, \ldots , 4$  and $k=1, \ldots , 5$, we construct a decision tree of type $k$ for the decision table $R_n$ by the greedy algorithm $\mathcal{E}$ described in \cite{ent} and find values  $h_{\mathcal{E}}^{(k)}(R_n)$, $L_{\mathcal{E}}^{(k)}(R_n)$, and $l_{\mathcal{E}}^{(k)}(R_n)$ -- see results in Tables \ref{tab6.3a}--\ref{tab6.5a}. Minimum values for each $n$ are in bold.
The obtained experimental results show that (i) for the depth, trees of types 2--5 outperform trees of type 1 for $n=2, \ldots , 4$,  (ii) for the number of
realizable nodes, trees of types 3 and 5 outperform trees of type 1 for $n=2$, and (iii) for the length of derived decision rules, trees of types 2--5 outperform trees of type 1 for $n=2, \ldots , 4$.

\begin{table}[h!]
\caption{Experimental results for greedy algorithm $\mathcal{A}_{\mathcal{E}}$ (depth)}
\label{tab6.3a}
\begin{tabular}{cccccc}
\toprule $n$ &  $h_{\mathcal{E}}^{(1)}(R_n)$ &  $h_{\mathcal{E}}^{(2)}(R_n)$ &  $h_{\mathcal{E}}^{(3)}(R_n)$ &  $h_{\mathcal{E}}^{(4)}(R_n)$ &  $h_{\mathcal{E}}^{(5)}(R_n) $ \\
\midrule
2 & 3 & \bf 2 & \bf 2 & \bf 2 & \bf 2 \\
3 & 6 & \bf 3 & \bf 3 & \bf 3 & \bf 3 \\
4 & 10 & \bf 6 & \bf 6 & \bf 6 & \bf 6 \\
\bottomrule &  &  &  &  &
\end{tabular}%
\end{table}

\begin{table}[h!]
\caption{Experimental results for greedy algorithm $\mathcal{A}_{\mathcal{E}}$ (number of realizable nodes)}
\label{tab6.4a}
\begin{tabular}{cccccc}
\toprule $n$ &  $L_{\mathcal{E}}^{(1)}(R_n)$ &  $L_{\mathcal{E}}^{(2)}(R_n)$ &  $L_{\mathcal{E}}^{(3)}(R_n)$ &  $L_{\mathcal{E}}^{(4)}(R_n)$ &  $L_{\mathcal{E}}^{(5)}(R_n) $ \\
\midrule
2 & 11 & 12 & \bf 9 & 12 & \bf 9 \\
3 & \bf 39 & 76 & 58 & 76 & 58 \\
4 & \bf 335 & 8,850 & 1,969 & 8,850 & 1,969 \\
\bottomrule &  &  &  &  &
\end{tabular}%
\end{table}

\begin{table}[h!]
\caption{Experimental results for greedy algorithm $\mathcal{A}_{\mathcal{E}}$
(length of derived rules)}
\label{tab6.5a}
\begin{tabular}{cccccc}
\toprule $n$ & $l_{\mathcal{E}}^{(1)}(R_n)$ & $l_{\mathcal{E}}^{(2)}(R_n)$ &
$l_{\mathcal{E}}^{(3)}(R_n)$ & $l_{\mathcal{E}}^{(4)}(R_n)$ & $l_{\mathcal{E}}^{(5)}(R_n) $ \\
\midrule
2 & 2.67 & \bf 2.17 & 2.33 &
\bf 2.17 & 2.33\\
3 & 4.55 & 3.50 & \bf 3.45 &
3.50 & \bf 3.45\\
4 & 7.65 & \bf 5.58 & 5.94 &
\bf 5.58 & 5.94\\
\bottomrule & & & & &
\end{tabular}%
\end{table}

\section{Sorting}
\label{S6.3}

In this section, we study the problem of sorting $n$ pairwise different elements from linearly ordered set, $n=3, \ldots, 6$.
Let $x_1, \ldots ,x_n$ be pairwise different elements from a linearly ordered set. We should find a permutation $(p_1, \ldots ,p_n)$ from the set $P_n$ of all permutations of the set $\{1, \ldots ,n\}$ for which $x_{p_1}<\cdots <x_{p_n}$. To this end, we use attributes $s_{i,j}$ such that $i,j \in \{1, \ldots ,n\}$, $i <j$, $s_{i,j} =1$ if $x_i<x_j$, and $s_{i,j} =0$ if $x_i>x_j$.

The problem of sorting $n$ elements can be represented as a decision table $S_n$ with $n(n-1)/2$ conditional attributes $s_{i,j}$,  $i,j \in \{1, \ldots ,n\}$, $i <j$, and $n!$ rows corresponding to permutations from $P_n$. For each permutation $(p_1, \ldots ,p_n)$, the corresponding  row of $S_n$ is labeled with this permutation as the decision. This row is filled with the values of the attributes  $s_{i,j}$ such that $s_{i,j}=1$ if and only if $i$ comes before $j$ in the tuple  $(p_1, \ldots ,p_n)$. The table $S_n$ can also be considered as a representation of an exact learning problem \cite{Angluin04}: conditional attributes form a domain, each row describes a concept, and rows form a concept class.
The decision table $S_3$ is shown in Fig. \ref{fig3}. 

\begin{figure}[h]
	\centering
$S_{3}=\;$%
\begin{tabular}{|ccc|c|}
\hline
$s_{1,2}$ & $s_{1,3}$ & $s_{2,3}$  & \\
\hline
1 & 1 & 1 &  $(1, 2, 3)$\\
1 & 1 & 0 &  $(1, 3, 2)$\\
0 & 1 & 1 &  $(2, 1, 3)$\\
0 & 0 & 1 &  $(2, 3, 1)$\\
1 & 0 & 0 &  $(3, 1, 2)$\\
0 & 0 & 0 &  $(3, 2, 1)$\\
\hline
\end{tabular}
\caption{Decision table $S_3$}
		\label{fig3}
\end{figure}

For $n=3, \ldots , 6$  and $k=1, \ldots , 5$, we find values of $h^{(k)}(S_n)$ and $L^{(k)}(S_n)$ using dynamic programming algorithms $\mathcal{A}_{h}$ and $\mathcal{A}_{L}$ described in \cite{ele} -- see results in Tables \ref{tab6.1} and \ref{tab6.2}. Minimum values for each $n$ are in bold.
The obtained experimental results show that the minimum depth of decision trees of  types 2--5 is one less than the minimum depth of decision trees of  type 1. The minimum number of realizable nodes in decision trees of types 3 and 5  is noticeably less than the minimum number  of realizable nodes in decision trees of  type 1. Decision trees of  types 2 and 4 have too many nodes.

\begin{table}[h!]
    \caption{Experimental results for dynamic programming algorithm $\mathcal{A}_{h}$}
\label{tab6.1}
\begin{tabular}{cccccc}
\toprule
$n$  &  $h^{(1)}(S_n)$ & $h^{(2)}(S_n)$ & $h^{(3)}(S_n)$ & $h^{(4)}(S_n)$ & $h^{(5)}(S_n)$ \\ \midrule
3 &	3 &	\bf 2 &	\bf 2 &	\bf 2 &	\bf 2 \\
4 &	5 &	\bf 4 &	\bf 4 &	\bf 4 &	\bf 4 \\
5 &	7 &	\bf 6 &	\bf 6 & \bf 6 &	\bf 6 \\
6 &	10 &	\bf 9 &	\bf 9 &	\bf 9 &	\bf 9 \\
 \bottomrule
\end{tabular}
\end{table}

\begin{table}[h!]
    \caption{Experimental results for dynamic programming algorithm $\mathcal{A}_{L}$}
\label{tab6.2}
\begin{tabular}{cccccc}
\toprule
$n$  &  $L^{(1)}(S_n)$ & $L^{(2)}(S_n)$ & $L^{(3)}(S_n)$ & $L^{(4)}(S_n)$ & $L^{(5)}(S_n)$ \\ \midrule
3 &	11 &	13 &	\bf 9 &	14 &	\bf 9 \\
4 &	47 &	253 &	\bf 39 &	254 &	\bf 39 \\
5 &	239 &	15,071 &	\bf 199 &	15,142 &	\bf 199 \\
6 &	1,439 &	2,885,086 &	\bf 1,199 &	2,886,752 &	\bf 1,199 \\
 \bottomrule
\end{tabular}
\end{table}

For $n=3, \ldots , 6$  and $k=1, \ldots , 5$, we construct a decision tree of type $k$ for the decision table $S_n$ by the greedy algorithm $\mathcal{E}$ described in \cite{ent} and find values  $h_{\mathcal{E}}^{(k)}(S_n)$, $L_{\mathcal{E}}^{(k)}(S_n)$, and $l_{\mathcal{E}}^{(k)}(S_n)$ -- see results in Tables \ref{tab6.3}--\ref{tab6.5}.  Minimum values for each $n$ are in bold.
The obtained experimental results show that (i) for the depth, trees of types 2--5 outperform trees of type 1 for $n=3, \ldots , 6$,  (ii) for the number of
realizable nodes, trees of types 3 and 5 outperform trees of type 1 for $n=3, 4$, and (iii) for the length of derived decision rules, trees of types 2--5 (especially trees of types 2 and 4) outperform trees of type 1 for $n=3, \ldots , 6$.

\begin{table}[h!]
\caption{Experimental results for greedy algorithm $\mathcal{E}$ (depth)}
\label{tab6.3}
\begin{tabular}{cccccc}
\toprule $n$ &  $h_{\mathcal{E}}^{(1)}(S_n)$ &  $h_{\mathcal{E}}^{(2)}(S_n)$ &  $h_{\mathcal{E}}^{(3)}(S_n)$ &  $h_{\mathcal{E}}^{(4)}(S_n)$ &  $h_{\mathcal{E}}^{(5)}(S_n) $ \\
\midrule
3       & 3    & \bf 2    & \bf 2    & \bf 2    & \bf 2    \\
4       & 5    & \bf 4    & \bf 4    & \bf 4    & \bf 4    \\
5       & 7    & \bf 6    & \bf 6    & \bf 6    & \bf 6    \\
6       & 10   & \bf 9    & \bf 9    & \bf 9    & \bf 9    \\
\bottomrule &  &  &  &  &
\end{tabular}%
\end{table}

\begin{table}[h!]
\caption{Experimental results for greedy algorithm $\mathcal{E}$ (number of realizable nodes)}
\label{tab6.4}
\begin{tabular}{cccccc}
\toprule $n$ &  $L_{\mathcal{E}}^{(1)}(S_n)$ &  $L_{\mathcal{E}}^{(2)}(S_n)$ &  $L_{\mathcal{E}}^{(3)}(S_n)$ &  $L_{\mathcal{E}}^{(4)}(S_n)$ &  $L_{\mathcal{E}}^{(5)}(S_n) $ \\
\midrule
3 & 11   & 14      & \bf 9    & 14      & \bf 9    \\
4 & 47   & 254     & \bf 39   & 254     & \bf 39   \\
5 & \bf 239  & 15,142   & 455  & 15,142   & 455  \\
6 & \bf 1,439 & 2,898,512 & 7,231 & 2,898,512 & 7,231 \\
\bottomrule &  &  &  &  &
\end{tabular}%
\end{table}

\begin{table}[h!]
\caption{Experimental results for greedy algorithm $\mathcal{E}$
(length of derived rules)}
\label{tab6.5}
\begin{tabular}{cccccc}
\toprule $n$ & $l_{\mathcal{E}}^{(1)}(S_n)$ & $l_{\mathcal{E}}^{(2)}(S_n)$ &
$l_{\mathcal{E}}^{(3)}(S_n)$ & $l_{\mathcal{E}}^{(4)}(S_n)$ & $l_{\mathcal{E}}^{(5)}(S_n) $ \\
\midrule
3 & 2.67 & \textbf{2.17} & 2.33 & \textbf{2.17} & 2.33\\
4 & 4.67 & \textbf{3.13} & 4.33 & \textbf{3.13} & 4.33\\
5 & 6.93 & \textbf{4.05} & 6.13 & \textbf{4.05} & 6.13\\
6 & 9.58 & \textbf{5.01} & 7.96 & \textbf{5.01} & 7.96\\
\bottomrule & & & & &
\end{tabular}%
\end{table}

\section{Conclusions}
\label{S6.4}

In this paper, we studied two problems: the problem of recognition of monotone Boolean functions with $n$ variables, $n=2, \ldots, 4$, and the problem of sorting $n$ pairwise different elements from linearly ordered set, $n=3, \ldots, 6$. For each problem, we found the minimum depth and the minimum number of realizable nodes in decision trees solving the problem.
We also found the depth and the number of realizable nodes in decision trees  constructed by entropy-based greedy algorithm and study the length of decision rules derived from these trees.

\bibliographystyle{splncs04}
\bibliography{two_problems}
\end{document}